# Colossal dielectric constants in single-crystalline and ceramic CaCu$_3$Ti$_4$O$_{12}$ investigated by broadband dielectric spectroscopy

S. Krohns, P. Lunkenheimer[a)]
*Experimental Physics V, Center for Electronic Correlations and Magnetism, University of Augsburg, 86135 Augsburg, Germany*

S. G. Ebbinghaus
*Solid State Chemistry, University of Augsburg, 86135 Augsburg, Germany*

A. Loidl
*Experimental Physics V, Center for Electronic Correlations and Magnetism, University of Augsburg, 86135 Augsburg, Germany*

In the present work the authors report results of broadband dielectric spectroscopy on various samples of CaCu$_3$Ti$_4$O$_{12}$, including so far only rarely investigated single crystalline material. The measurements extend up to 1.3 GHz, covering more than nine frequency decades. We address the question of the origin of the colossal dielectric constants and of the relaxational behavior in this material, including the second relaxation reported in several recent works. For this purpose, the dependence of the temperature- and frequency-dependent dielectric properties on different tempering and surface treatments of the samples and on ac-field amplitude are investigated. Broadband spectra of a single crystal are analyzed by an equivalent circuit description, assuming two highly resistive layers in series to the bulk. Good fits could be achieved, including the second relaxation, which also shows up in single crystals. The temperature- and frequency-dependent intrinsic conductivity of CCTO is consistent with the Variable Range Hopping model. The second relaxation is sensitive to surface treatment and, in contrast to the main relaxation, also is strongly affected by the applied ac voltage. Concerning the origin of the two insulating layers, we discuss a completely surface-related mechanism assuming the formation of a metal-insulator diode and a combination of surface and internal barriers.

## I. INTRODUCTION

For further technical advances in the performance and miniaturization of capacitive electronic elements, new materials with high dielectric constant ($\varepsilon'$) are urgently needed. Thus, since the first reports of extremely high ("colossal") values of $\varepsilon'$ in CaCu$_3$Ti$_4$O$_{12}$ (CCTO),[1,2,3] this material is in the focus of scientific research (see, e.g., Refs. 4,5,6,7,8,9) and more than 180 papers were published on this topic in the last six years. However, while in the early stages various intrinsic mechanisms were proposed,[2,3,10] further investigations revealed strong hints that the colossal $\varepsilon'$ in CCTO is of non-intrinsic origin.[4,5,6,7,9,11] In a pioneering work, Sinclair *et al.*[4] plotted the results of their dielectric measurements on ceramic CCTO in the complex impedance plane representation, revealing a succession of two semicircles. Such a behavior is often found in ceramic materials, whose dielectric response at low frequencies is dominated by grain boundaries. Hence these results were explained with an "internal barrier layer capacitor" (IBLC) model: Polarization is built up at insulating grain boundaries between the semiconducting grains of the ceramic samples. This generates non-intrinsic colossal values of the dielectric constant, accompanied by a strong Maxwell-Wagner (MW) relaxation mode. However, it should be noted that the impedance representation does not allow identifying the physical origin of the observed semicircles: The first semicircle could be due to grain boundaries, planar crystal defects (e.g., twinning boundaries), insulating surface layers, or even an intrinsic relaxation. Indeed, based on measurements of ceramic CCTO samples, significant contributions from external surface effects were proposed by our group[9] as also detected by us in various other materials.[8,12,13] We found that sputtered contacts give rise to a much higher $\varepsilon'$ than silver paint contacts and that $\varepsilon'$ depends on sample thickness. Very recently, we reported similar behavior also for single crystalline samples.[14] These results can be explained within a "surface barrier layer capacitor" (SBLC) picture assuming, e.g., the formation of a Schottky diode at the contact-bulk interface.[9]

Nevertheless, it is frequently stated in current literature that the IBLC is the more likely mechanism. To large extent, this notion is based on results on ceramic samples with different grain sizes, prepared, e.g., by subjecting the samples to various heat treatments. This was shown to result in a strong variation of the absolute values of $\varepsilon'$ and conductivity $\sigma'$, thus supporting the SBLC picture (e.g., Refs. 11,15,16). However, it should be noted that extremely high values of $\varepsilon'$ particularly were observed in CCTO single crystals (SCs),[3] where grain boundaries are absent. Thus other internal boundaries in SCs were considered, e.g., twin boundaries.[4,6,17] However, if there are any planar defects in SCs that generate high dielectric

---
[a)]Electronic mail: peter.lunkenheimer@physik.uni-augsburg.de

constants and strong relaxations, they may be expected to also contribute to a separate relaxational response in polycrystals (PCs), where often the grains are rather large (up to 100 μm). In this case, in PCs two relaxations should be observed, one from planar defects and the other one originating from the grain boundaries. Interestingly, there is indeed evidence for at least two relaxations in CCTO. Already in one of the earliest reports on CCTO[2] indications for a second relaxation were obtained and since then various experiments revealed that it is quite a common feature of polycrystalline CCTO.[9,10,14,15,18,19] This relaxation results in even larger $\varepsilon'$ values than the famous main relaxation, albeit at lower frequencies and/or higher temperatures only. But, in contrast to the above-mentioned scenario where the two relaxations should arise from grain boundaries and planar defects within the grains, also an alternative explanation is possible. Namely, one relaxation could be due to an IBLC and the other to a SBLC mechanism,[9,14] a notion that would provide a solution of the SBLC/IBLC debate.

Obviously, there are many open questions concerning the origin of the dielectric properties of CCTO. In the present work we contribute to their solution by providing dielectric results on various samples that were subjected to different tempering and surface treatments. The dielectric constant, loss, and conductivity were determined in a broad frequency range of nine frequency decades, up to 1.3 GHz, which is of major relevance for possible high-frequency applications. The measurements also include single-crystalline samples, which so far only were rarely studied.[3] The present work extends our previous study on single- and polycrystalline CCTO where we reported evidence for a second relaxation also in SCs and a strong influence of surface treatments on the dielectric properties.[14] Here we provide additional experimental data and a more detailed analysis. Amongst others we have performed fits of the obtained spectra using an equivalent circuit approach and show data for samples subjected to different tempering treatments. In addition, also the ac voltage dependence of dielectric-constant and conductivity spectra is shown.

## II. EXPERIMENTAL DETAILS

Polycrystalline samples of CCTO were prepared as reported in Ref. 9 and sintered at 1000°C in air for up to 48 h. SCs were grown by the floating zone technique.[3,14] The applied growth furnace (model GERO SPO) is equipped with two 1000 W halogen lamps, the radiation of which is focused by gold-coated ellipsoidal mirrors. Polycrystalline bars serving as seed and feed rods were cold-pressed and sintered in air for 12 h at 1000°C. The seed rod was rotated with a speed of 30 rpm, while the feed was kept still. The growth rate was adjusted to 5 mm/h. To avoid thermal reduction of copper, crystal growth was performed in oxygen (flow rate 0.2 l/min) at a pressure of 4 bar. The surface topographies of the samples were investigated using an environmental scanning electron microscope (ESEM XL30, FEI Company) with an accelerating voltage of 15 kV. Operating in a low-hydrogen-pressure atmosphere prevents charging effects without contacting the sample surface.

For the dielectric measurements silver paint or sputtered gold contacts (thickness 100 nm) were applied to opposite sides of the plate-like samples. Some samples were subjected to different sintering and surface treatments as described in the next section. All sintering procedures were performed without any contacts applied on the samples to avoid diffusion effects of the contact material. The complex conductivity $\sigma^* = \sigma' + i\sigma''$ and permittivity $\varepsilon^* = \varepsilon' - i\varepsilon''$ were measured over a frequency range of up to nine decades (1 Hz < $\nu$ < 1.3 GHz) at temperatures down to 40 K as detailed in Refs. 9 and 20. In most experiments the applied ac voltage was 1 V. In addition also measurements up to 15 V were performed. The geometry and typical grain sizes of the samples, for which results are shown in the present work, are given in Table I. The experimental results for samples treated in the same way (same thermal history, same contact preparation, same surface treatment) were highly reproducible.

| Sample | shown in Fig. | area (mm$^2$) | thickness (mm) | grain size (μm) |
|---|---|---|---|---|
| SC | 1,2,3,4 | 16.6 | 1.10 | not applicable |
| PC 1, 3h tempered | 5 | 24.0 | 1.24 | 2-6 |
| PC 2, 48h tempered | 5 | 32.6 | 1.20 | 80-200 |
| PC 3, 48h tempered, before polishing | 6(a,d) | 22.9 | 1.63 | 80-200 |
| PC 3, 48h tempered, after polishing | 6(b,e), 6(c,f), 7 | 22.9 | 1.55 | 80-200 |

TABLE I. Sample parameters. The grain sizes are rough estimates only, obtained from ESEM measurements.

## III. RESULTS AND DISCUSSION

### A. Temperature dependence

Fig. 1(a) shows the temperature dependence of the dielectric constant and conductivity of single crystalline CCTO for various frequencies. This representation corresponds to that used in Fig. 2(a) of the pioneering work of Homes et al.[3] Here we provide data on one SC, measured with two different types of contacts (silver paint and sputtered gold). In both cases, at low frequencies and high temperatures colossal values of the dielectric constant show up. With $\varepsilon'_{high} \approx 2\times10^5$ they are significantly higher for the sputtered than for the silver paint contacts. This agrees with our findings from frequency-dependent data on PCs[9] and SC,[14] which were interpreted assuming a surface-related origin of the high $\varepsilon'$ values.[9] The well-known relaxation mode of CCTO shows up as a step like decrease of $\varepsilon'$ with decreasing temperature, which shifts to higher temperatures with increasing frequency. It is now generally accepted that this relaxation is of MW type, caused by an equivalent circuit consisting of the bulk contribution, connected in series to a parallel RC circuit with R and C much higher than the corresponding bulk quantities. The highly resistive thin layers that generate this RC circuit could be surface layers, grain boundaries, or planar defects.

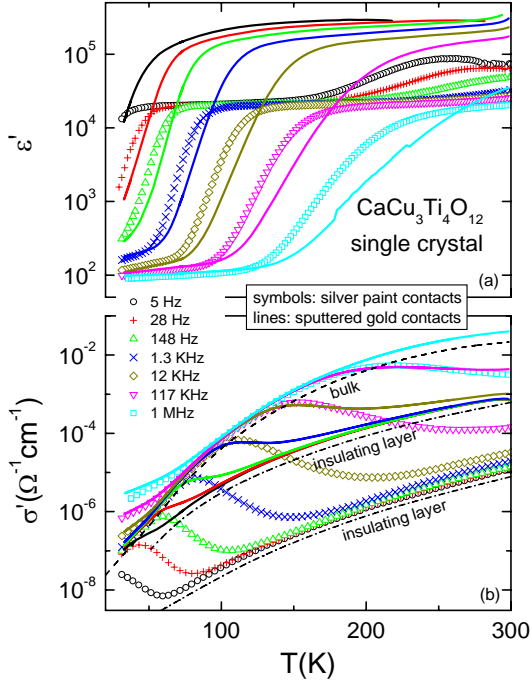

FIG. 1. Temperature-dependent dielectric constant (a) and conductivity (b) of single-crystalline CCTO with sputtered gold (solid lines) and silver paint contacts (symbols) at various frequencies (lines and symbols of same frequency have same color). The dashed and dash-dotted lines in (b) give an estimate of the intrinsic bulk dc-conductivity (same for both measurements) and the contribution of the insulating layer (different), respectively. For clarity reasons these lines were slightly shifted downwards.

Interestingly, in the $\varepsilon'(T)$ results obtained with silver-paint contacts, a second relaxation mode is detected. It leads to an increase of the dielectric constant, e.g., at $T > 180$ K for 5 Hz, which is superimposed to the high-$\varepsilon'$ plateau of the main relaxation mode. This finding is consistent with the detection of a second relaxation in the dielectric spectra of single-crystalline CCTO reported by us in Ref. 14 (see also Fig. 3(a)). The occurrence of a second relaxation also in SCs excludes one of the scenarios mentioned in section I: As the effect of grain boundaries in single crystals is expected to be negligible, it is not possible to explain the two relaxations in CCTO by separate contributions from grain boundaries and planar defects.

The conductivity, shown in Fig. 1(b) is directly related to the dielectric loss $\varepsilon''$ via $\sigma' = 2\pi\nu\varepsilon''\varepsilon_0$ ($\varepsilon_0$ the permittivity of free space). Thus its temperature dependence is identical to that of the loss and it exhibits the well-known peaks and their characteristic shift with frequency. However, in contrast to the loss representation,[3] in the conductivity representation the peaks merge at low temperatures. This is the case not only for different frequencies of a single sample preparation but also for the measurements with different contact types shown in Fig. 1(b). This behavior is due to the fact, that at the low-temperature wing of these peaks, the condition $\nu > 1/(2\pi\tau)$ is fulfilled, with $\tau$ the relaxation time of the equivalent circuit. Thus here the capacitances of the insulating layers are shorted and the intrinsic dc-conductivity of CCTO is detected. Its temperature dependence is indicated by the uppermost dashed line in Fig. 1(b) (for better readability, it is slightly shifted downwards). The deviations from this line showing up, e.g., below about 70 K for 1 MHz, point to a frequency dependence of the intrinsic $\sigma'$ due to hopping charge transport.[21]

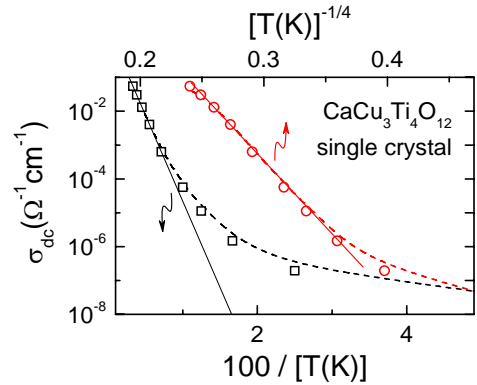

FIG. 2. (Color online) Dashed lines: Temperature dependence of the dc conductivity approximated by the dashed line shown in Fig. 1(b). In addition, results for $\sigma_{dc}$ from the fits of the frequency dependence (Fig. 3) are shown (symbols). Both data sets are provided in two representations that should lead to a linearization for Arrhenius and VRH behavior indicated by the solid lines.

In Fig. 2, we show the bulk dc conductivity, approximated by the dashed line of Fig. 1(b), in an Arrhenius representation (dashed line, lower scale). In addition, $\sigma_{dc}$ results from the fits of frequency-dependent data discussed below are included as squares. As indicated by the solid line, only in a rather restricted range at high temperatures the data follow thermally activated Arrhenius behavior.[22] As an alternative, the same data are also shown in a representation that leads to linear behavior for 3D-Variable Range Hopping (VRH, upper scale).[23] Indeed, in a large temperature region the data follow the prediction of this model. VRH behavior in CCTO was also reported in Ref. 24, however, based on data in a smaller temperature range only. Within this model, a frequency dependence of the conductivity, $\sigma' = \sigma_{dc} + \sigma_0 \omega^s$ ($s < 1$) is predicted. This explains the mentioned intrinsic frequency dependence of $\sigma'$ below roughly 70 K (Fig. 1(b)).

Also at high temperatures the $\sigma'(T)$ curves merge, however, approaching two different curves for the two contact types (dash-dotted lines). Here the conductivity is dominated by the highly resistive layer and obviously its resistance differed by about two decades for the two types of contact preparations. In Ref. 9, the very different values of $\varepsilon'_{high}$ obtained for sputtered and silver-paint contacts were interpreted assuming a surface-related origin of the high $\varepsilon'$ values via formation of Schottky diodes at the contact-bulk interface. It was assumed that different surface wetting of these contact types, arising from larger air gaps at the interface for silver-paint contacts,[8,25] should lead to less effective formation of the Schottky diodes.[8] One can speculate that this may also explain the very different high-temperature behavior of the conductivity observed in Fig. 1(b). An alternative explanation of the marked differences in the dielectric properties of CCTO measured with different contact materials could be that due to their grain structure and the air gaps, silver-paint contacts are generally unsuited for the measurements of very high dielectric constant materials. However, we demonstrated in two earlier works[9,26] that for intrinsic $\varepsilon'$ values up to $2\times10^4$ and $2\times10^5$, respectively, silver paint contacts lead to the same results as sputtered gold. Even the intrinsic values of $\varepsilon'$ exceeding $10^7$ observed in o-TaS$_3$, a typical charge density wave system,[27] could be properly measured using silver paint contacts.[28]

## B. Frequency dependence

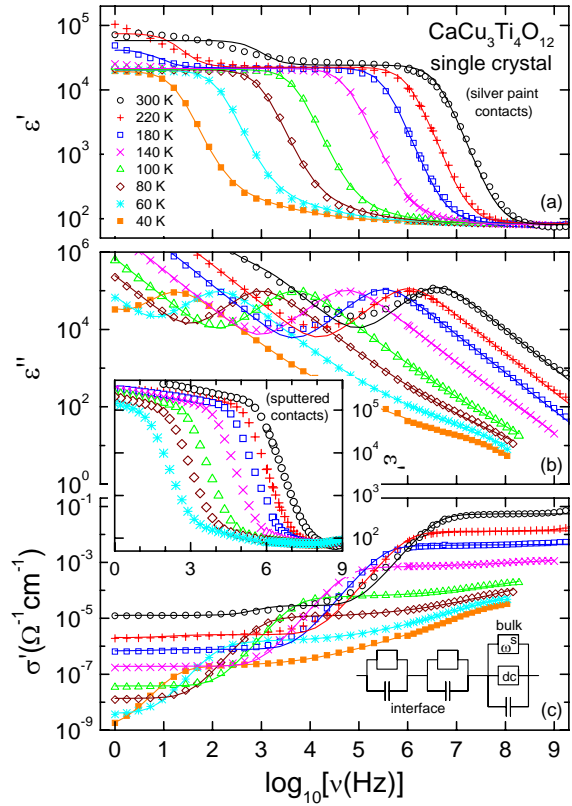

FIG. 3. (Color online) Frequency-dependent dielectric constant (a), loss (b), and conductivity (c) of single-crystalline CCTO with silver paint contacts at various temperatures. The data is fitted with the equivalent circuit indicated in (c).[9] The inset shows $\varepsilon'$ for the same sample using sputtered gold contacts.[14]

Figure 3 shows broadband spectra of dielectric constant, loss, and conductivity for single crystalline CCTO with contacts formed by silver paint. The most obvious differences to the corresponding figure for sputtered contacts[14] are the lower values of the dielectric constant $\varepsilon'_{high}$ (cf. inset) and the clear occurrence of a second relaxation at low frequencies. Both features are also revealed in the temperature-dependent data (Fig. 1). The MW relaxation modes show up as steplike decreases in $\varepsilon'(\nu)$, shifting through the frequency window with temperature. Correspondingly, peaks in the loss $\varepsilon''(\nu)$ and shoulders in $\sigma'(\nu)$ are revealed. Considering the above-mentioned equivalent circuit, at high frequencies the highly resistive thin layers (surface layers, grain boundaries, or planar defects) that generate the RC circuit in series to the bulk, are shorted via their capacitance and the intrinsic behavior is detected. This is nicely demonstrated in Fig. 4, where the frequency-dependent dielectric constant and conductivity for the two contact types are compared. While at low frequencies strong differences show up, the intrinsic properties of CCTO governing the high-frequency behavior remain unaffected of the contact variation. They reveal a rather high dielectric constant of about 85 and a succession of dc and ac conductivity, characteristic for hopping of localized charge carriers.[9,29] The latter is consistent with our finding of VRH from the dc conductivity (Fig. 2).[21,23]

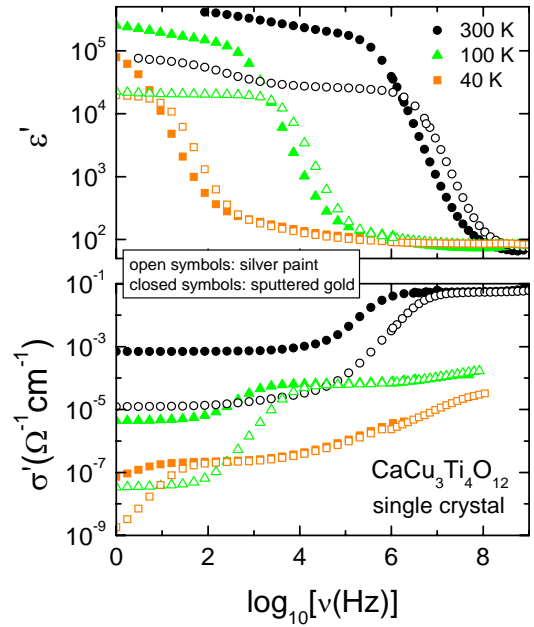

FIG. 4. (Color online) Frequency-dependent dielectric constant (a) and conductivity (b) of single-crystalline CCTO with silver paint contacts (open symbols) and sputtered gold (closed symbols) at selected temperatures.

To account for the second relaxation, an additional RC element in the equivalent circuit has to be assumed. Adding a further element for the mentioned frequency-dependent conductivity, we arrive at the circuit shown in Fig. 3(c), which also was used in Ref. 9 for the analysis of the behavior of PCs. The lines in Fig. 3 are least-square fits with this model, leading to reasonable agreement of experimental data and fits. Only the second relaxation in $\varepsilon'(\nu)$ is not perfectly covered: The experimental relaxation steps are more smeared out than the fit curves. This points to a distribution of relaxation times of the circuit due to heterogeneities. Depending on the interpretation of this relaxation they may be generated, e.g., by surface roughness or a distribution of grain sizes. In Figs. 1 and 3, while the relaxation steps in $\varepsilon'$ are accompanied by corresponding features also in $\varepsilon''$ and $\sigma'$ (a peak and a shoulder, respectively), this seems not to be the case for the second relaxation. However, taking a closer look at Fig. 3(c), a small shoulder, e.g., at about $10^3$ Hz for 300 K is observed. This becomes especially obvious in the fits, which slightly over exaggerate this feature. If we ascribe the two $RC$ circuits to two different types of insulating layers, the sum of their resistances dominates the conductivity at the lowest frequencies. For the 300 K curve in Fig. 3(c), for $\nu > 10^3$ Hz one circuit becomes shorted, which leads to the small shoulder and the corresponding relaxation step in $\varepsilon'$ (Fig. 3(a)). Obviously, the resistances of both layers are of similar magnitude, which explains the small effect in $\sigma'$ and the directly related $\varepsilon''$. This is corroborated by the results from the equivalent circuit analysis: With values of 14.4 nF and 4.2 nF at 300 K, the capacitances of the two layers differ more than the resistances (37.6 kΩ and 14.6 kΩ, respectively). As expected, the bulk capacitance (10.7 pF) and resistance (12.2 Ω) are much smaller than those of the layers. In Fig. 2, the temperature-dependent conductivity values calculated from the bulk resistances obtained for different temperatures are shown. They agree well with the intrinsic dc conductivity estimated from the temperature-dependent plot of Fig. 1(b).

It is well known that the dielectric properties of CCTO PCs are qualitatively similar to that of SCs but usually they reveal a lower range of the dielectric constant $\varepsilon'_{high}$. However, as recently shown, when using sputtered contacts and subjecting the samples to long sintering procedures, values up to $10^5$ can be reached also in PCs.[14] To systematically study the effect of different tempering times, we used silver-paint contacts, which, in contrast to sputtered contacts, can be easily removed by dissolving. Figure 5 shows two typical results of our sintering experiments: $\varepsilon'_{high}$ increases strongly when tempering the sample for 48 h. A sample tempered for 24 h (not shown) revealed $\varepsilon'_{high}$ values between those of the 3 h and 48 h tempered sample. In the limit of high frequencies and low temperatures, similar intrinsic $\varepsilon'$ and $\sigma'$ values are reached, as expected for MW relaxations. Within the IBLC framework, the strong tempering effect on $\varepsilon'_{high}$ can be ascribed to the growth of grain sizes.[11,15,16] However, as an alternative it can also be explained in terms of the different surface topographies revealed by ESEM (insets of Fig. 5). When the grains are growing during sintering, obviously the surface of the samples becomes much smoother. Thus the effect of tempering could be quite the same as that of using sputtered contacts instead of silver paint: Just in the way as the smaller particles of sputtered contacts lead to an increase of the direct contact area between the rough sample surface and the metallic contact material,[8,9] the smoother sample surface also could result in a better "wetting" of the contact.[25] If one assumes the formation of a Schottky diode leading to an insulating depletion layer at the contact to the semiconducting CCTO sample, this better wetting should cause an increase of $\varepsilon'_{high}$ as indeed observed in Fig. 5. Overall, as the tempering obviously also has an effect on the sample surface, it is difficult to draw definite conclusions from such experiments.

### C. Second relaxation

In all figures shown so far, at low frequencies and/or high temperatures there are at least indications of the mentioned second relaxation. As for the main relaxation, also for this relaxation three possible origins have to be considered: intrinsic, IBLC, or a surface layer. An intrinsic mechanism is highly unlikely as the magnitude of this relaxation varies considerably for different samples (see present work and Refs. 2,9,10,14,15,18,30). In Ref. 30 an intrinsic relaxor ferroelectric behavior was proposed, based on the occurrence of broad dielectric peaks in $\varepsilon'(T)$ at low frequencies and high temperatures. In fact, in the present work such temperature dependence is observed, too: It is revealed by the peak-like behavior of $\varepsilon'(T)$ at 5 Hz in Fig. 1(a) as well as the crossing of the curves for 220 K and 300 K at low frequencies in Fig. 3(a). However, such behavior can also arise within an equivalent circuit description with temperature independent capacitors (to keep it simple, in the following we assume only a single $RC$ circuit in series to the bulk). For example, making the reasonable assumption that the insulating-layer capacitance $C_i$ is much larger than the bulk capacitance $C_b$, the overall capacitance in the limit of small frequencies is given by $C_0 = C_i G_b^2/(G_b+G_i)^2$ ($G_b$ and $G_i$ denote the bulk and insulating-layer conductance, respectively). For $G_b \gg G_i$, $C_0 \approx C_i$ is obtained, but if $G_b(T)$ and $G_i(T)$ should become of similar magnitude at high temperatures, $C_0$ can decrease to $C_i/4$. Together with the decrease of $\varepsilon'$ with decreasing temperature due to the MW relaxation, this can lead to a peak. Thus the temperature dependence of $G_b$ and $G_i$ can result in a maximum in $C_0$, and thus also in the low-frequency $\varepsilon'(T)$, as observed in Ref. 30 and Fig. 1(a).

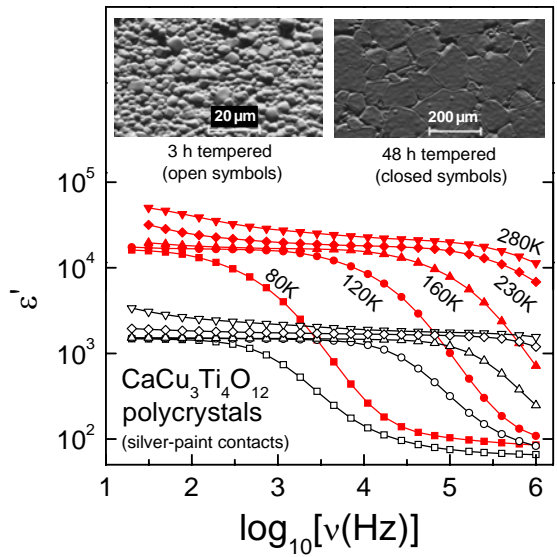

FIG. 5. (Color online) Frequency-dependent dielectric constant of ceramic CCTO samples tempered for 3 h (open symbols) and 48 h (closed symbols) with silver paint contacts. The insets show the surface topographies obtained by ESEM.

In our search for the origin of the second relaxation, we found a strong effect when removing the surface layers of a PC by polishing.[31] The polishing reduced the original sample thickness of 1.63 mm by about 0.08 mm (Tab. I). Judging from our thickness-dependent experiments in Ref. 9, the mere thickness variation should lead to a change of $\varepsilon'$ not exceeding 5%. To take account of the possible presence of a surface layer with different stoichiometry, the polishing was done under $N_2$ atmosphere to avoid any reaction with oxygen at the freshly exposed surface. Afterwards the sample was installed in the vacuum of the cryostat as fast as possible. Fig. 6 shows the results for the untreated sample (a, d), immediately after polishing (b, e), and after leaving it in air for 48 h (c, f). In the dielectric constant of the untreated sample (Fig. 6(a)), the second relaxation is only weakly seen, at best. After polishing it becomes very prominent (Fig. 6(b)). It again seems to weaken when exposed to air (with removed silver paint contacts) and a shift of the relaxation steps to lower frequencies is observed (Fig. 6(c)). The conductivity plotted in Figs. 6(d)-(f) also shows some significant and interesting effects: The intrinsic properties remain unaffected; e.g., the bulk dc-conductivity, indicated by the horizontal dashed lines, is identical for all three cases. However, the downward step in $\sigma'(\nu)$, preceding the dc plateau, reveals strong differences: Its lower plateau value, seen at room temperature at the lowest frequencies only, is much smaller for the untreated sample. For the treated samples an intermediate plateau shows up (e.g., for 140 K at $\sigma' \approx 10^{-5}$ $\Omega^{-1}$cm$^{-1}$ and $\nu \approx 10^3$ Hz). After 48 h exposure to air, it becomes significantly reduced. Considering the equivalent circuit shown in Fig. 1(c), the first plateau at lowest frequencies (seen at room temperature only) is determined by the resistance of the first RC circuit, the intermediate plateau by that of the second circuit, and the third one by the bulk conductivity.

The results of Fig. 6 seem to indicate that the second relaxation is influenced by surface effects and exchange with atmosphere. Thus, based on the experimental results of the present and earlier works of our group,[8,9,14] it is suggestive to assume a surface-generated origin of both relaxations in CCTO. Then the second relaxation should be caused by a second surface layer. A possible explanation of such a layer arises from the fact that CCTO, like other transition metal oxides, strongly tends to oxygen understoichiometry developing during preparation.[4] This leads to the introduction of charge carriers and an increase of $\sigma'$. However, one may speculate that at the surface of the samples an exchange with ambient atmosphere leads to a layer closer to ideal stoichiometry and thus more insulating than the bulk.[19] But there are also other possibilities: For example chemical decomposition into more simple oxides could occur at the surface, also leading to a thin insulating layer. Within these scenarios, the polishing either has removed the surface layer only partly, reducing its thickness and thus increasing the apparent static $\varepsilon'$ of the second relaxation, or it was completely removed. In the latter case the high $\varepsilon'$ should be due to the development of a very thin new layer in the time when the sample was exposed to air during insertion into the cryostat. However, it is not clear if this could happen during this rather short time of about 3 min. Another explanation is that in $N_2$ atmosphere decomposition can occur at the surface of the sample.[19] In any case, the fact that the relaxation varies with time when the sample is left in air (Figs. 6(c) and (f)) demonstrates that the second relaxation is sensitive to exchange with the surrounding atmosphere. This excludes other scenarios as, e.g., possible deterioration of crystal structure at the surface by polishing, as it is known from Si wafers. In the picture of a surface layer generated by different oxygen-stoichiometry, the effect of air-exposure leads to a growth of the thickness of the insulating layer; thus the capacitance decreases as observed at low frequencies in Figs. 6(b) and (c).

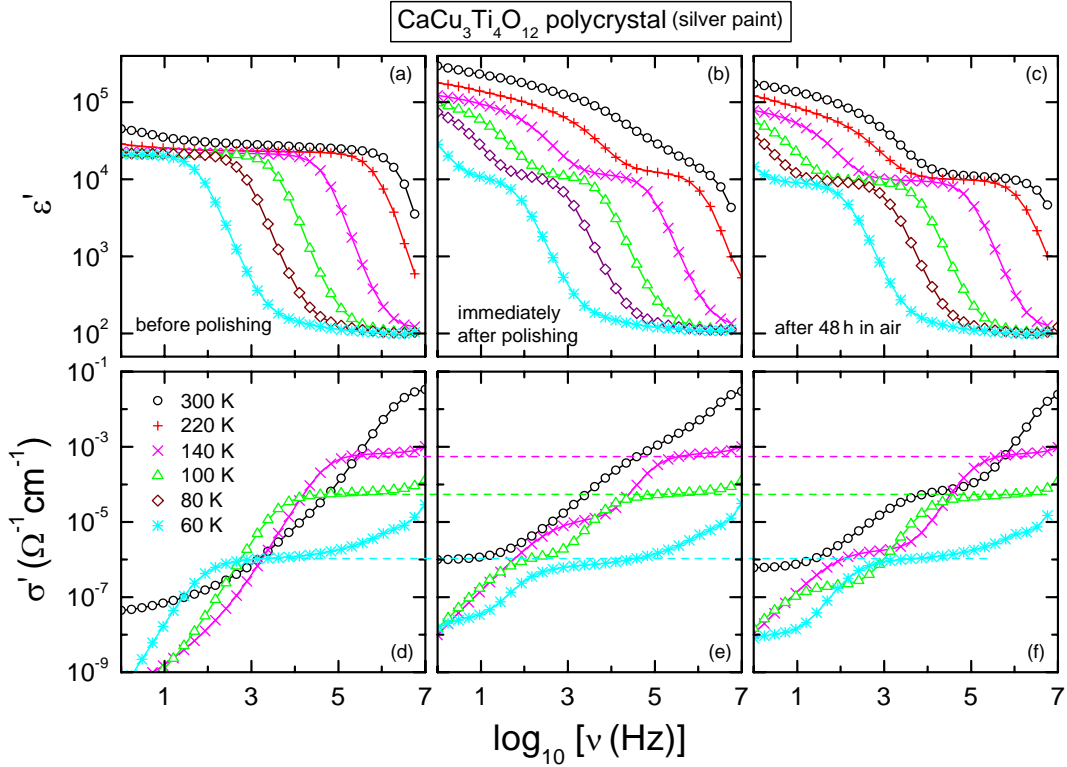

FIG. 6. (Color online) Frequency-dependent dielectric constant (a-c) and conductivity (d-f) of polycrystalline CCTO (tempered 48 h) with silver paint contacts at various temperatures. Results are shown for the untreated sample (a, d), after polishing under $N_2$ atmosphere (b, e; both contacts reapplied) and after leaving it in air (without silver paint contacts) for 48 h (c, f).

If indeed there is an insulating layer of different stoichiometry at the surface of the samples, earlier arguments invoking the formation of Schottky diodes at the contact/bulk interfaces are no longer valid. Instead we would have the situation of a metal covering a thin insulating layer on top of a semiconductor. This should lead to the generation of a so-called metal-insulator-semiconductor (MIS) diode, well known from electronic textbooks and forming a part of field effect transistors. In this scenario, a depletion layer forms just as in the case of a conventional diode, however situated at the surface of the bulk semiconductor, just below the insulating layer. As for a conventional diode, a strong effect of different wetting can be expected. This model implies the following succession of layers (from the inside): semiconducting bulk / depletion layer with lower conductance / insulating chemically different layer with lowest conductance / metal. In $\sigma'(\nu)$, with increasing frequency one thus should see a transition from the very low conductivity arising from the insulating non-stoichiometric layer, to the higher conductivity related to the depletion layer, followed by another steplike increase to the intrinsic bulk conductivity. Such a behavior is indeed revealed in Figs. 6(e) and (f). However, within this framework it is not straightforward to explain the variation of the absolute values of the conductivity, especially those of the above-mentioned intermediate plateau. It is likely that the variation of this plateau value, via the relaxation time of the equivalent circuit, leads to the strong frequency shift of the second relaxation when the sample is subjected to air for 48 h (compare Figs. 6(b) and (c)). One even could speculate that for the untreated sample the intermediate plateau is several decades lower (of the order of $10^{-7}$ for room temperature) and thus the second relaxation is located below the lowest measuring frequency and of similar magnitude as for the treated samples. Thus it is not clear if indeed the thickness of the different-stoichiometric surface layer or the conductance of the depletion layer, i.e. the effectiveness of the depletion, changes in Fig. 6.

### D. Voltage dependence

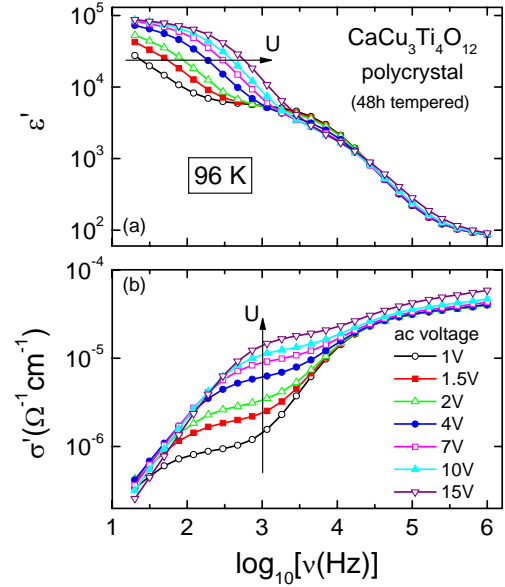

FIG. 7. (Color online) $\varepsilon'(\nu)$ (a) and $\sigma'(\nu)$ (b) at 96 K for different ac voltages. The same ceramic sample as in Fig. 6(c) and (f) was investigated.

If any kind of diode formation plays a role in the generation of the dielectric behavior of CCTO, a non-trivial dependence of the dielectric behavior on the ac voltage should result. Thus dielectric spectra of the 48 h tempered polished PC, where both relaxations are well developed [cf. Fig. 6(c)], were measured for different voltages at 96 K. The results, shown in Fig. 7, reveal a strong dependence of the frequency of the second relaxation step on the ac voltage, while the main relaxation remains practically unaffected. Also the intermediate plateau value of $\sigma'(\nu)$ is strongly voltage dependent. To explain these findings, let us consider the equivalent circuit shown in Fig. 1. We make the assumption that $C_b \ll C_1 \ll C_2$ and $R_b \ll R_1 \ll R_2$ (the indices 1 and 2 denote the elements of the two $RC$ circuits and "b" the intrinsic bulk behavior). Then the relaxation time of the main relaxation is approximately given by $\tau_{main} \approx R_b C_1$. The capacitance $C_1$ corresponds to the plateau value $\varepsilon'_{high}$ of the main relaxation (e.g., $\varepsilon'_{high} \approx 5 \times 10^3$ in Fig. 7(a)). Obviously, it is not or only weakly dependent on voltage. It also is plausible that the bulk conductivity, which can be read off above about $3 \times 10^4$ Hz in Fig. 7(b), is independent of voltage. This explains the absence of any significant variation of the main relaxation step in Fig. 7(a). The relaxation time of the second, low-frequency relaxation can be approximated by $\tau_{second} \approx R_1 C_2$. Judging from Fig. 7(a), $C_2$ leading to the low-frequency plateau of $\varepsilon'(\nu)$ (about $10^5$), seems to be constant. However, $R_1$, leading to the intermediate plateau in $\sigma'(\nu)$ between $10^2$ Hz and $10^4$ Hz in Fig. 7(b) is strongly voltage dependent. This explains the voltage-dependent shift of the second relaxation observed in Fig. 7(a).

The voltage dependence of $R_1$ may well reflect the non-linear I-V characteristics of a diode. This is consistent with the MIS-diode framework mentioned above. Within this picture, $C_2$ corresponds to the capacitance of the insulating surface layer, which can be expected to be insensitive to voltage, as indeed observed. However, in contrast to this completely SBLC dominated picture, the results of Fig. 7 also can be explained assuming a combination of an SBLC and IBLC mechanism: Its is clear that the second relaxation is generated by the second $RC$-circuit composed of $R_2$ and $C_2$. However, the variation of its relaxation time that leads to the observed shift in Fig. 7(a) completely is due to a variation of the first $RC$ circuit, namely of $R_1$. Thus Fig. 7 only does imply that the first relaxation caused by $R_1$ and $C_1$ arises from a diode (the origin of which is still to be clarified)), but this cannot be said for the second relaxation. Instead, the latter could be due to the planar defects mentioned in section 1. Then we could have Schottky diodes at the metal-bulk interface causing the first relaxation and planar defects (not grain boundaries) causing the second relaxation. Even the results of Fig. 6 may be explainable in this way: As mentioned above, the strong differences in the appearance of the second relaxation caused by the different surface treatments could be predominantly ascribed to a variation of its relaxation time. In this way, the observed surface sensitivity of the second relaxation could be due to the variation of $R_1$ of the first $RC$ circuit, i.e. the contact resistance. Within this model, it is this resistance that corresponds to the above-mentioned strongly varying intermediate plateau in Figs. 6(d)-(f). However, it is not clear, why the polishing and air exposure should lead to such a strong variation, while the capacitance $C_1$ is only weakly affected (see plateau values of $\varepsilon'_{\text{high}} \approx 10^4$ in Figs. 6(a)–(c)). In addition, a close inspection of Fig. 6 also seems to reveal an amplitude variation of the second relaxation, which cannot be explained in this way.

## IV. SUMMARY AND CONCLUSIONS

In summary, we performed detailed dielectric measurements on various CCTO samples subjected to different surface and heat treatments. Our study includes so far only rarely investigated single crystalline samples and covers a frequency range up to 1.3 GHz. We found a strong variation of $\varepsilon'$ with contact material also in SCs of CCTO. Besides the well-known main relaxation, indications for a second relaxation are found in most of our measurements, including SCs. This leads to the conclusion that there must be two different types of insulating layers in CCTO. Indeed, using an equivalent circuit that takes account of this notion, reasonable fits of the spectra extending over nine frequency decades could be achieved. Its origin so far is unclarified and it also is not clear if its variing visibility in different samples is due to a variation of its amplitude or relaxation time. As we detected the second relaxation also in single crystals, grain boundaries may not play any role. We found that its relaxation time and amplitude depends sensitively on surface treatment[14] and ac voltage.

There are two alternative explanations of our experimental findings. For the first one, we consider the presence of a thin surface layer with different stoichiometry, leading to the formation of a MIS diode at the sample surface. Within such a completely surface-dominated picture we can explain most of our own results, but clearly this is at variance with the numerous reports evidencing an IBLC mechanism found in literature. While we have shown that tempering experiments lead to simultaneous variation of grain boundaries and surface and thus may be less significant, experiments like the microcontact measurements reported in Ref. 7 are difficult, if not impossible to explain within an SBLC framework. As an alternative to the developed surface dominated picture, a combination of contributions of planar defects (causing the second relaxation) and contact-generated Schottky diodes (causing the main relaxation) also seems to be consistent with most of our data. While the latter explanation has some problems as mentioned at the end of the preceding section, we have to state clearly that from the current experimental basis it is not possible to definitely conclude, which one of the above developed scenarios is correct. However, irrespective of any model description, the bare experimental fact that both relaxations in CCTO vary when the surface of the sample is modified, be it by contact variation or by polishing, shows that surface-related effects at least play some role for the dielectric properties of CCTO. In any case, we do not claim the complete absence of any grain boundary contributions in our or other ceramic samples. Even within our surface-dominated picture, these contributions could arise outside or at the edge of the frequency/temperature window or be superimposed by the surface contributions.

Finally, we want to point out that, concerning its intrinsic properties, CCTO is unusual mainly in one way, namely its rather high $\varepsilon'$ of about 85, arising from a high ionic polarizability.[22] The colossal $\varepsilon'$ of this material is directly related to this high value as it can be assumed that the dielectric constant of the insulating layers (e.g., planar defect or surface depletion layer) is related to that of the bulk and mainly the number of charge carriers is different. This high intrinsic $\varepsilon'$, together with the small thickness of these layers (whatever their origin may be) leads to the apparently colossal $\varepsilon'$. Concerning possible applications of CCTO in modern microelectronic, one faces two major problems: At first the dielectric loss of CCTO is too high. In addition, as found by us for SCs (Fig. 3) and PCs,[9] the values of $\varepsilon'$ at frequencies beyond MHz, which is a technically highly relevant regime, are far from being colossal. Within the equivalent circuit framework, in order to shift the main relaxation towards higher frequencies, the bulk resistance should be lowered. Thus if looking for possible alternative materials, one should aim at lower bulk resistivity and an as high or higher intrinsic $\varepsilon'$ as CCTO. To ensure a low overall loss, the resistance of the highly resistive thin layers should be as high as possible. Thus it is essential to identify the mechanisms leading to these layers. We believe that the presented results at least may provide some hints that will help solving this question.

**ACKNOWLEDGMENTS**

This work was supported by the Commission of the European Communities, STREP: NUOTO, NMP3-CT-2006-032644 and by the DFG via the SFB 484.

[1] M. A. Subramanian, D. Li, N. Duan, B. A. Reisner, and A. W. Sleight, J. Solid State. Chem. **151**, 323 (2000).
[2] A. P. Ramirez, M. A. Subramanian, M. Gardel, G. Blumberg, D. Li, T. Vogt, and S. M. Shapiro, Solid State Commun. **115**, 217 (2000).
[3] C. C. Homes, T. Vogt, S. M. Shapiro, S. Wakimoto, and A. P. Ramirez, Science **293**, 673 (2001).
[4] D. C. Sinclair, T. B. Adams, F. D. Morrison, and A. R. West, Appl. Phys. Lett. **80**, 2153 (2002).
[5] M. A. Subramanian and A. W. Sleight, Solid State Sci. **4**, 347 (2002); L. He, J. B. Neaton, M. H. Cohen, D. Vanderbilt, and C. C. Homes, Phys. Rev. B **65**, 214112 (2002).
[6] M. H. Cohen, J. B. Neaton, L. He, and D. Vanderbilt, J. Appl. Phys. **94**, 3299 (2003)
[7] S.-Y. Chung, I.-D. Kim, and S.-J. L. Kang, Nature Mat. **3**, 774 (2004).
[8] P. Lunkenheimer, V. Bobnar, A. V. Pronin, A. I. Ritus, A. A. Volkov, and A. Loidl, Phys. Rev. B **66**, 052105 (2002).
[9] P. Lunkenheimer, R. Fichtl, S. G. Ebbinghaus, and A. Loidl, Phys. Rev. B **70**, 172102 (2004).
[10] A. P. Ramirez, G. Lawes, V. Butko, M. A. Subramanian, and C. M. Varma, cond-mat/0209498.
[11] T. B. Adams, D. C. Sinclair, and A. R. West, Adv. Mater. **14**, 1321 (2002); B. A. Bender and M.-J. Pan, Mat. Sci. Eng. B **117**, 339 (2005).
[12] P. Lunkenheimer, M. Resch, A. Loidl, and Y. Hidaka, Phys. Rev. Lett. **69**, 498 (1992); A. I. Ritus, A. V. Pronin, A. A. Volkov, P. Lunkenheimer, A. Loidl, A. S. Shcheulin, and A. I. Ryskin, Phys. Rev. B **65**, 165209 (2002); V. Bobnar, P. Lunkenheimer, J. Hemberger, A. Loidl, F. Lichtenberg, and J. Mannhart, Phys. Rev. B **65**, 155115 (2002); P. Lunkenheimer, T. Rudolf, J. Hemberger, A. Pimenov, S. Tachos, F. Lichtenberg, and A. Loidl, Phys. Rev. B **68**, 245108 (2003).
[13] B. Renner, P. Lunkenheimer, M. Schetter, A. Loidl, A. Reller, and S. G. Ebbinghaus, J. Appl. Phys. **96**, 4400 (2004); P. Lunkenheimer, T. Götzfried, R. Fichtl, S. Weber, T. Rudolf, A. Loidl, A. Reller, and S. G. Ebbinghaus, J. Solid State Chem. **179**, 3965 (2006).
[14] S. Krohns, P. Lunkenheimer, S. G. Ebbinghaus, A. Loidl, Appl. Phys. Lett. **91**, 022910 (2007); *ibid*. **91**, 149902 (2007).
[15] J. L. Zhang, P. Zheng, C. L. Wang, M. L. Zhao, J. C. Li, and J. F. Wang, Appl. Phys. Lett. **87**, 142901 (2005).
[16] G. Zang, J. Zhang, P. Zheng, J. Wang and C. Wang, J. Phys. D: Appl. Phys. **38**, 1824 (2005); B. S. Prakash and K. B. R. Varma, Physica B **382**, 312 (2006); L. Ni, X. M. Chen, X. Q. Liu and R. Z. Hou, Solid State Commun. **139**, 45 (2006).
[17] N. Kolev, R. P. Bontchev, A.J. Jacobson, V. N. Popov, V. G. Hadjiev, A. P. Litvinchuk, and M. N. Iliev, Phys. Rev. B **66**, 132102 (2002); L. Wu, Y. Zhu, S. Park, S. Shapiro, G. Shirane, and J. Tafto, Phys. Rev. B **71**, 014118 (2005); J. Li, A. W. Sleight, and M. A. Subramanian, Solid State Commun. **135**, 260 (2005); M.-H. Whangbo and M. A. Subramanian, Chem. Mater. **18**, 3257 (2006).
[18] L. Fang, M. Shen, and W. Cao, J. Appl. Phys. **95**, 6483 (2004).
[19] C. C. Wang and L. W. Zhang, Appl. Phys. Lett. **88**, 042906 (2006).
[20] R. Böhmer, M. Maglione, P. Lunkenheimer, and A. Loidl, J. Appl. Phys. **65**, 901 (1989); U. Schneider, P. Lunkenheimer, A. Pimenov, R. Brand, Ferroelectrics **249**, 89 (2001).
[21] S. R. Elliott, Adv. Phys. **36**, 135 (1987); A. R. Long, Adv. Phys. **31**, 553 (1982).
[22] Ch. Kant, T. Rudolf, F. Mayr, S. Krohns, P. Lunkenheimer, S. G. Ebbinghaus, and A. Loidl, arXiv:0709.1065 (unpublished)
[23] N. F. Mott and E. A. Davis, *Electronic Processes in Non-Crystalline Materials* (Clarendon Press, Oxford, 1979).
[24] L. Zhang and Z.-J. Tang, Phys. Rev. B **70**, 174306 (2004).
[25] J. H. Hwang, K. S. Kirkpatrick, T. O. Mason, E. J. Garboczi, Solid State Ionics **98**, 93 (1997).
[26] D. Starešinić, P. Lunkenheimer, J. Hemberger, K. Biljaković, and A. Loidl, Phys. Rev. Lett. **96**, 046402 (2006).
[27] D. Starešinić, K. Biljaković, W. Brütting, K. Hossein, P. Monceau, H. Berger, and F. Levy, Phys. Rev. B **65**, 165109 (2002).
[28] D. Starešinić, private communication.
[29] A. Tselev, C. M. Brooks, S. M. Anlage, H. Zheng, L. Salamanca-Riba, R. Ramesh, and M. A. Subramanian, Phys. Rev. B **70**, 144101 (2004).
[30] S. Ke, H. Huang, and H. Fan, Appl. Phys. Lett. **89**, 182904 (2006).
[31] A similar experiment was performed in Ref. 19, however a complete vanishing of the second relaxation was claimed only after reducing the sample thickness by 0.82 mm, which would imply too thick a layer to arrive at colossal values of *ε'*. Possibly reoxidation occurred during polishing in air.